\documentclass[]{aa}
\usepackage{natbib}
\usepackage{psfig}
\usepackage{txfonts}
\def\AV{\mbox{A$_{\rm V}$}}

\def\HH{\mbox{H$_2$}}

\def\Xe{\mbox{${\rm X(e)}$}}

\def\nH2{\mbox{${\rm n}_\HH}$}

\def\NH2{{\rm N}({\rm H}_2)}
\def\pccc{~{\rm cm}^{-3}} 
\def\ccc{~{\rm cm}^3} 
\def\pcc {~{\rm cm}^{-2}}

\def\Tstar#1 {\mbox{${\rm T}_{\rm #1}^*$}}
\def\Tsub#1 {\mbox{${\rm T}_{\rm #1}$}}
\def\TK  {\Tsub K }
\def\TB  {\Tsub B }

\def\Texc {\Tsub ex }

\def\p{\mbox{$^+$}}
\def\cotw {\mbox{$^{12}$CO}}
\def\coth {\mbox{$^{13}$CO}}
\def\coei {\mbox{C$^{18}$O}}

\def\hcop{\mbox{{HCO\p}}} 
\def\hTWcop{\mbox{{H\cotw\p}}} 
\def\hTHcop{\mbox{{H\coth\p}}}
\def\TWCp{\mbox{{$^{12}$C\p}}} 
\def\THCp{\mbox{{$^{13}$C\p}}} 
\def\TWC{\mbox{{$^{12}$C}}} 
\def\THC{\mbox{{$^{13}$C}}} 
\def\WCO{\mbox{W$_{\rm CO}$}}

\def\h13cop{\mbox{{H$^{13}$CO\p}}}

\def\c3h2{\mbox{C$_3$H$_2$}}
 
 \def\R0{R$_0$}

\def\ddeg{{}^\circ\kern-.1em}

\def\kms{\mbox{km\,s$^{-1}$}}
\def\ps{\mbox{s$^{-1}$}}

\def\E#1 {$10^{#1}$}
\def\E#1 {E{#1}}
\def\P#1,{$\nH2\TK~=~#1\times~10^4\pccc$~K}
\def\ec#1,#2,#3,{#1\,(#2)\E{#3}}

\def\zoph{$\zeta$ Oph}

\def\H3{\mbox{H$_3$}}

\def\zetaH{\mbox{$\zeta_H$}}
\def\RH2{\mbox{R$_{\rm G}$}}
\def\fH2{\mbox{f$_{\HH}$}}
\def\FH2{\mbox{F$_{\HH}$}}

\def\g13{\mbox{g$_{13}$}} 

\title{Formation, fractionation and excitation of carbon monoxide in diffuse clouds}
\author{H. S. Liszt\inst{1} }
\institute{National Radio Astronomy Observatory,
           520 Edgemont Road,
           Charlottesville, VA,
           USA 22903-2475}

\begin{document}
\date{received \today}
\offprints{H. S. Liszt}
\mail{hliszt@nrao.edu}
%
\abstract
   {A wealth of observations of CO in absorption in diffuse
   clouds has accumulated in the past decade at $uv$ and mm-wavelengths}
{Our aims are threefold: a) To compare the  $uv$ and mm-wave results; 
  b) to interpret  \coth\ and \cotw\ abundances in terms of the physical
   processes which separately and jointly determine them; c) to interpret 
 observed J=1-0 rotational excitation and line brightness in 
 terms of ambient gas properties.}
{A simple phenomenological model of CO formation as the immediate 
    descendant of quiescently-recombining \hcop\ is used to study the 
    accumulation, fractionation and rotational excitation of CO in more 
    explicit and detailed models of \HH-bearing diffuse/H I clouds}
{The variation of N(CO) with N(\HH) is explained by quiescent 
  recombination of a steady fraction n(\hcop)/n(\HH) = $2\times 10^{-9}$.
  Observed N(\cotw))/N(\coth) ratios generally do not require a special 
  chemistry but result from competing processes and do not 
   provide much insight into the local gas properties, especially the 
  temperature. J=1-0 CO line brightnesses directly represent N(CO), 
  not N(\HH), so the CO-\HH\ conversion factor varies widely; it 
  attains typical values at N(\cotw) $\la 10^{16}\pcc$.  
   Models of CO rotational excitation account for the line 
  brightnesses and CO-\HH conversion factors but readily reproduce 
  the observed excitation temperatures 
  and optical depths of the rotational transitions only if excitation 
  by H-atoms is weak -- as seems to be the case for the very most recent
  calculations of these excitation rates.}
{ Mm-wave and $uv$ results generally agree well but the former 
  show somewhat more enhancement of \THC\ in \coth . In any case, 
  fractionation may seriously bias \TWC/\THC\ ratios measured in 
  CO and other co-spatial molecules.  Complete C$\rightarrow$CO conversion
  must occur over a very narrow range of \AV\ and N(\HH) just beyond
  the diffuse regime.  For  N(\HH) $< 7\times10^{19}\pcc$ the character
  of the chemistry changes inasmuch as CH is generally undetected while
  CO suffers no such break.  }
\keywords{ interstellar medium -- molecules }

\maketitle

\section {Introduction.}

Except for hydrogen, carbon monoxide is the most important and widely
observed molecule in the interstellar medium (ISM).  The 7-8 decade 
span in column density over which CO is directly observed, from 
N(CO) $= 10^{12}\pcc$ in {\it uv} absorption in the diffuse interstellar 
medium to N(CO) $> 10^{19}\pcc$ in mm and sub-mm emission from dense 
and giant molecular clouds, is exceeded only by that of \HH\ itself.
The ubiquity of CO has encouraged the use of mm-wave CO emission as 
a possible tracer of molecular hydrogen even into such extreme 
environments as high velocity clouds \citep{DesCom+07} .

Interpreting observations of CO in diffuse gas \citep{SnoMcC07} 
over the lower half of its range, at N(CO) $\la 10^{16}\pcc$, has 
been particularly challenging.  The fraction of free gas-phase 
carbon in CO is small, 
a few percent or less  when, locally, \AV $<$ 1 mag, but it is still 
30-50 times larger than can be explained by the quiescent gas-phase 
ion-molecule chemistry of low-density media like diffuse clouds
\citep{vDiBla88,WarBen+96}.  
The relative abundance of CO with respect to \HH\ varies widely
in this regime (see Fig. 1; with much scatter, approximately as 
N(CO) $\propto$ N(\HH)$^2$ over the range 
X(CO) = N(CO)/N(\HH) $ \approx 3\times 10^{-8} - 3\times 10^{-5}$) and 
the relative abundances of \cotw\ and \coth\ are strongly affected 
by fractionation \citep{WatAni+76,SmiAda80} such that 
$20 \la $ N(\cotw)/N(\coth) $\la 170$ (see the references cited in 
Sect. 2)

Despite the comparatively small CO abundances in diffuse clouds, 
$\lambda$ 2.6mm J=1-0 rotational emission is often appreciable.
Typically it is seen that $\TB \approx 1-5$ K, 
\WCO $\approx 1-5$ K \kms\ peak or integrated brightness for 
N(CO) $= 10^{15}-10^{16}\pcc$ \citep{LisLuc98} but peak brightnesses 
as high as 10-13 K have been observed \cite{LisLuc94}. Moreover, the 
J=1-0 rotational transition may have appreciable optical depth in 
diffuse gas because its excitation is quite weak. Surveys of 
\cotw\ alone may be hard-pressed to distinguish between 
dark and diffuse gas, especially at higher galactic latititude or
larger galactocentric radii owing to the broader distribution
of diffuse gas.


To elucidate the properties of CO in the diffuse regime, we discuss 
here a wealth of observational material at {\it uv} and mm-wavlengths 
which has accumulated (much of it very recently) over the past decade.
The plan of this work is as follows.  Section 2 gathers the 
previously-published observational results \citep{LisLuc98,
SonWel+07,BurFra+07,SheRog+07} which form the basis of the 
present discussion.  Section 3 displays and discusses
the run of observed values of the CO and \HH\ column densities
to demonstrate that there is at least a phenomenological basis
for understanding the abundance of CO in diffuse gas, in order
to show that there is some knowledge of the microscopic CO formation 
rate. This rate and those of the various other physical processes
which account for the abundances of \cotw\ and \coth\ are set out 
in detail in Sect. 4.  Section 5 discusses the observed abundances
of \cotw\ and \coth\ and their fractionation, and Sect. 6 
discusses the rotational excitation and brightness of the
J=1-0 rotational transition.  Section 7 is a brief discussion and 
summary of outstanding concerns.

\section{Observational material}

With one minor exception (see Sect. 6) , the discussion here relies 
on previously-published results in the {\it uv} and mm-wavelength regimes, 
as we now discuss.

\subsection{{\it uv} and optical absorption}

Many determinations of N(CO) and N(\HH) have recently been published 
by \cite{SonWel+07}, \cite{BurFra+07} and 
\cite{SheRog+07}, who provide measurements of N(\HH), N(\cotw) and
N(\coth), along with such additional physically interesting quantities 
as the J=1-0 rotational temperatures of \HH\ and/or carbon monoxide 
and, from \cite{SonWel+07}, column densities and related quantities 
for such species as CH and C$_2$.  For the lines of sight 
where there is overlap, agreement is generally excellent for the column
densities of CO and \HH\ and only slightly worse for the CO excitation
temperature, as shown below in Figs. 5 and 6.  In cases of overlap,
we chose values from the reference with the smaller quoted 
errors if the datasets  were equally comprehensive 
in that direction.  However, we also chose not to mix values for 
N(\cotw) and N(\coth) from different references along any given line 
of sight, because the systematic errors could be different.

\subsection{MM-wave absorption}

Observations of carbon monoxide in absorption toward mm-wave continuum
sources were given by \cite{LisLuc98}, along with rotational excitation
temperatures and isotope ratios, $etc.$).  N(\HH) is not known directly
in these measurements, which assume instead that  N(\HH) = 
N(\hcop)/$2\times10^{-9}$\ \citep{LisLuc96,LucLis96,LucLis00}.  The 
current discussion may be regarded as a consistency check on this assumption.

In comparing the radio and optical lines of sight, it should be remembered
that the former use extragalactic background sources, penetrate the 
entire galactic layer, and refer to individual, well-resolved kinematic 
components.  The optical/{\it uv} lines of sight stop within the Galaxy and 
are sums over undifferentiated -- though not necessarily 
blended -- features for species studied in the {\it uv}, that is for CO and \HH .

\begin{figure}
\psfig{figure=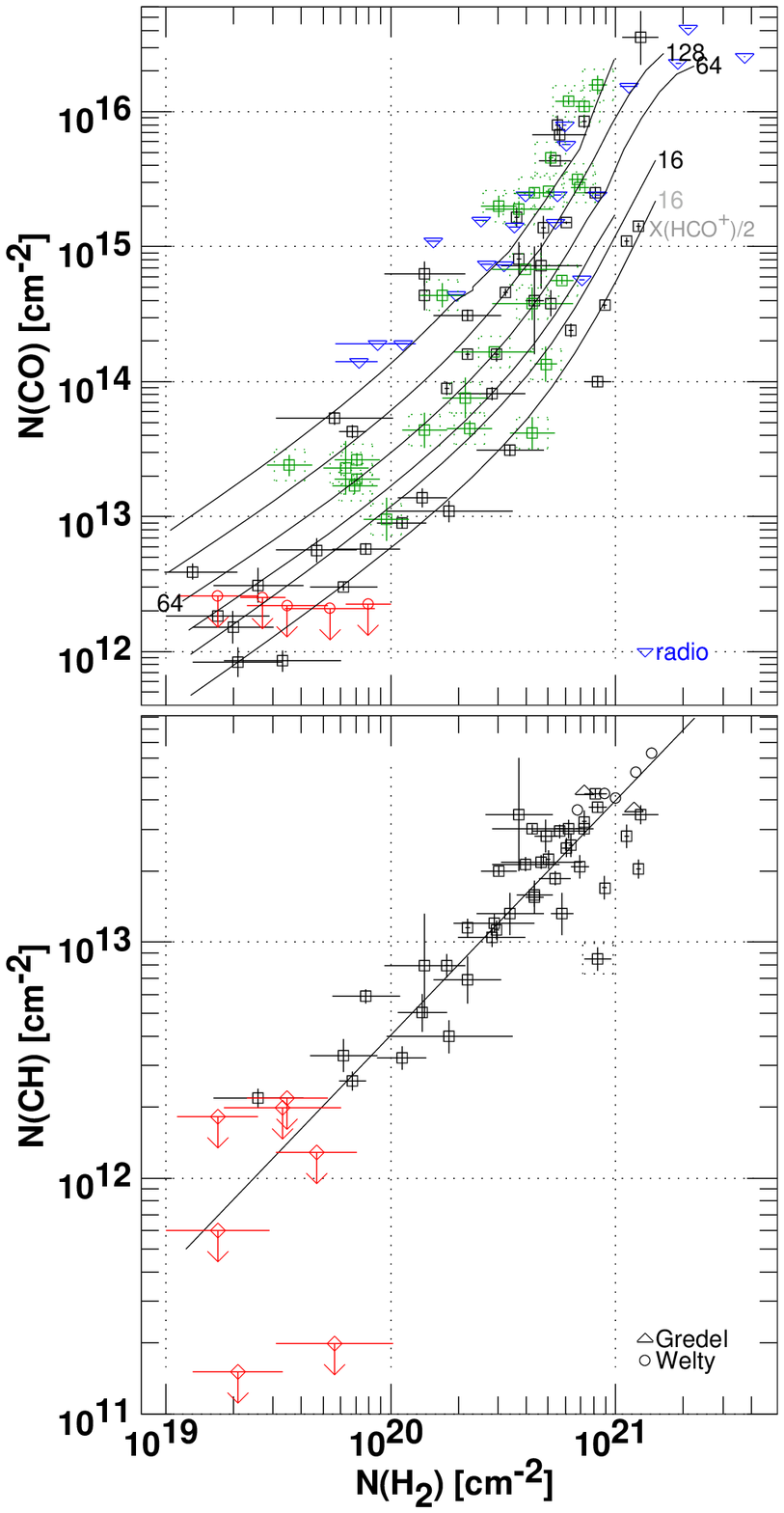,height=16cm}
\caption[]{Top:  variation of CO and \HH\ column densities 
 observed in {\it uv} (rectangles; \cite{SonWel+07} and 
 \cite{BurFra+07}) and mm-wave (triangles; \cite{LisLuc98}) CO 
 absorption.  Lines of sight represented in Fig. 5 are outlined
 and shown in green.  
For the mm-wave data N(\HH) = N(\hcop)/$2\times10^{-9}$. The curves
represent models of CO formation $via$ recombination of \hcop\ 
and are labelled by their density  n(H) $\approx$ n(H I) + 2n(\HH) 
(see Sect. 4). The lowest curve shows the result of halving the \hcop\ abundance
at n(H) $= 16 \pccc$. 
  Bottom: Variation of CH and \HH\ column densities from the summary tables
 of \cite{SonWel+07} and a few high-N(CH) datapoints from \cite{GreVan+93}
and \cite{WelHob+03}. }
\end{figure}

\section{Variation of molecular abundance with \HH\ and the 
source function for carbon monoxide}

The abundances of \cotw\ and \coth\ are the result of several
processes; chemical formation and destruction, carbon isotope exchange 
and (selective) photodissociation.  Interpreting the 
observations requires an understanding of the competing influences
of all such processes, but the overall guage is really set by
a comparison of the photodissociation rate (which is known in 
free space and calculable within a model) and the direct chemical 
formation rate.  Once the latter is specified, other physical 
properties follow directly and it is even possible to model the 
internal rotational excitation and predicted mm-wave brightness. 
Here we outline an empirical approach to estimating the rate of
CO formation.

\subsection{CO and \HH }

In Fig. 1 at top we show the observed run of N(\cotw) with N(\HH).
The rise of N(CO) with increasing N(\HH) is very steep, approximately
N(CO) $\propto $ N(\HH)$^2$, with scatter amounting to two orders of
magnitude at fixed N(\HH). At each N(\HH) the radio-derived N(CO) is
comparatively but not inordinately large; this could be an indication 
that X(\hcop), taken as $2\times 10^{-9}$, has been slightly 
overestimated.  The large subset of {\it uv}-absorption datapoints which 
have CO excitation temperature measurements and which therefore 
appear also in Fig. 5 have been flagged and colored green in Fig. 1,
in order to highlight the fact that most of them require model
densities of at least n(H) = 64 $\pccc$ (see Sect. 6).

What to make of the variation of N(CO) with N(\HH)?  Models of CO 
formation employing thermal processes in quiescent diffuse gas 
have largely been unable to reproduce observed 
values of X(CO) except perhaps at rather higher densities than are 
otherwise inferred for diffuse gas; see Fig. 16 of \cite{SonWel+07}.
However, purely phenomenologically, it is known that thermal 
gas-phase electron recombination of the observed amount of \hcop, 
X(\hcop) $\approx 2\times 10^{-9}$, does suffice to reproduce the 
observed X(CO) at modest densities.  This is shown by the curves
at the top in Fig. 1, where, updating the calculations of \cite{LisLuc00}, 
we plot the predicted N(CO) for diffuse cloud models in which an 
artificially steady relative abundance X(\hcop) $=2\times10^{-9}$ 
is allowed to recombine with free electrons at the thermal rate; 
note that the mm-wave observations of N(CO) and N(\hcop) have been 
placed in the plane of the Figure by relying on the same 
X(\hcop) which in the models explains the CO abundance.

The models are for small uniform density spheres immersed in the 
mean galactic radiation fields, within which the thermal 
balance,  ionization equilibrium
and \HH\ and CO accumulation problems are solved self-consistently 
\citep{Lis07} using the self-shielding factors of \cite{LeeHer+96}.
For each total density n(H) = 16, 32, 64 ... 256 $\pccc$, a series 
of models of increasing N(H) was calculated, and the CO and \HH\ 
column densities across the central line of sight were calculated
to form the curves for each n(H).
 
The newer calculations differ from the older ones in a variety of 
small ways, for instance the rates for 
O I and C I excitation by hydrogen atoms were very recently recalculated by 
\cite{AbrKre+07}.  Most important is the recognition, following 
observation of \H3\p, and accounting for the neutralization rate on 
small grains/PAH, that that the low-level hard ionization rate of 
hydrogen, presumably due to cosmic rays, is apparently 
larger in diffuse gas \citep{McCHin+02,Lis03}.  This has the effect of
increasing the densities of H\p\ and He\p\ and the overall electron
fraction  (which, however, is never more than about twice the free
carbon abundance).  Higher electron fractions lead to more rapid formation of CO 
(if CO forms by \hcop\ recombination) while the increased density of
He\p\ plays an elevated role in the destruction of carbon monoxide.
Indeed, destruction by He\p\ is the dominant {\it chemical} mechanism for
\cotw\ destruction.  {\it The various processes responsible for
destroying CO are discussed in Sect. 4.}

In the Figure, scatter in N(CO) at fixed N(\HH) is implicitly attributed 
to variations in density, which in the models ranges between 
n(H) = 16 $\pccc$ and 256 $\pccc$.  However, the very lowest curve
shows the result of halving X(\hcop) in the model and additional 
observational scatter , perhaps substantial, could also result from 
variations in geometry and ambient illumination.  Consideration of 
fractionation strongly suggests variations of a factor two or more in 
the photodissociation rate (see Sect. 5 and Fig. 4).

In the remainder of this work we will assume that the direct CO
formation rate is adequately specified by the dissociative recombination
of \hcop\ at X(\hcop) $= 2\times 10^{-9}$ and we will employ this 
as a basis to interpret the observed 
fractionation and rotational excitation in a self-consistent fashion 
(interpreting brightness, $etc.$ within the context of a model which
also accounts for the abundance).  The following Section sets out in 
detail the physical processes which determine the abundances of \cotw\ and
\coth.

\subsection{CO, CH and \HH }

By contrast with CO, the  variation of N(CH) with N(\HH) is 
very nearly linear and with considerably smaller scatter and 
overall range for the actually-measured N(CH).  
Fig. 1 at bottom is an updated version of Fig. 1 of \cite{LisLuc02} and 
although it shows a somewhat less-perfect correlation in a larger dataset 
(46 $vs$ 32 sightlines),  the derived means 
$<{\rm N(CH)}>/<{\rm N(\HH)}> =  4.1\times 10^{-8}$
and
$<{\rm N(CH)}/{\rm N(\HH)}> = 4.5\pm1.6\times10^{-8}$ (for lines
of sight with CH detections at N(\HH) $\ge 7\times 10^{19}\pcc$)
are nearly unchanged,  N(CH) and N(CO) are well-correlated 
(coefficient 0.86; this is somewhat contrary to a remark by 
\cite{CreFed04}) though the functional relationship is quite steep, 
N(CO) $\propto$ N(CH)$^{2.6}$.  

In general, for N(CH) $> 2\times 10^{12}\pcc$, CH is a quite reliable
indicator of the \HH\ column density, with  N(\HH) $\approx$ 
N(CH)$/4\times 10^{-8}$.
However, and somewhat remarkably, for N(\HH) $< 7\times 10^{19}\pcc$
it is somewhat more likely to find CO than CH, and several lines of
sight at lower N(\HH) have very high values of N(CO)/N(CH) (see Fig. 18 in
\cite{SonWel+07}).  Lines of sight with low molecular abundances,
including a few of those with smaller or undetected N(CH) in Fig. 1 were
discussed by \cite{ZsaFed03}, who argued that non-thermal processes
must preferentially dominate the chemistry along some very transparent 
lines of sight.  Alternatively we noted that very 
high values for X(OH) = N(OH)/N(\HH) might be  expected in 
more diffuse \HH-bearing gas just from quiescent thermal processes
 \citep{Lis07};  if CO is formed from OH, high N(CO)/N(CH) ratios
might also be explained in this way.

\section{Mechanisms of CO formation and fractionation}

In this Section we set out the various physical processes contributing to
the formation, destruction and fractionation of carbon monoxide, albeit
in a very reductive fashion.  The ambient gas is taken to be diffuse but 
purely molecular and the molecular abundances are taken to small enough 
that conservation of nuclei need not be explictly observed;  that is,
the fraction of C in CO is not large enough to affect the relative
abundance of \hcop\ and the fraction of $^{13}$C in \coth\ is not
large enough to alter the relative rates at which \cotw\ and \coth\
form from the isotopic variants of \hcop.

\begin{figure}
\psfig{figure=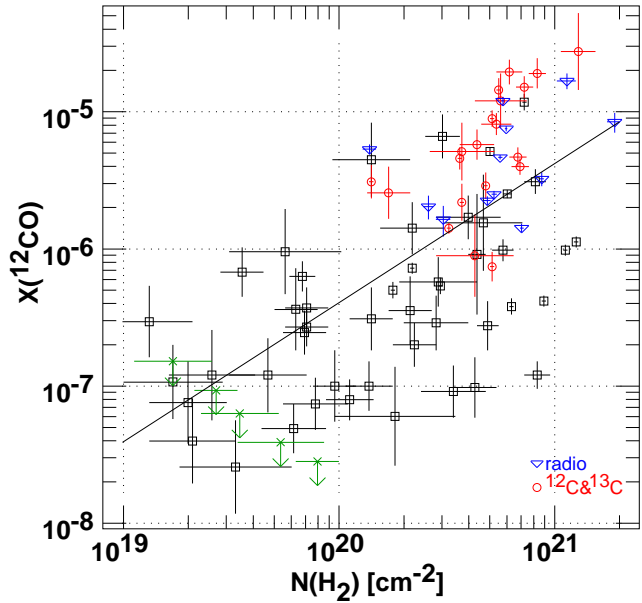,height=8cm}
\caption[]{Variation of X(\cotw)=N(\cotw)/N(\HH) with N(\HH).  
 The regression line shown has slope 1.014$\pm0.13$ and passes through X(\cotw)
 $= 4.05 \times 10^{-7}$ for N(\HH)$ = 10^{20}\pcc$ .  Lines of sight
having {\it uv} absorption measurements of both \cotw\ and \coth\ are noted 
(red in appropriate media); they have relatively high X(\cotw) at a 
given N(\HH).}
\end{figure}

\subsection{Source function for carbon monoxide formation}

As noted above, a source function for carbon monoxide may be approximated
as the quiescent thermal recombination rate of a fixed relative abundance 
X(\hcop) = $2\times10^{-9}$ such that the observations are explained at 
densities typical of diffuse clouds, even if the exact formation route for 
\hcop\ in diffuse gas is problematical.   Ion-molecule reactions 
in such a quiescent gas form too little \hcop\ and/or CO by a factor of 
about 30 and to account for the discrepancy it has been suggested that 
chemical reactions with C\p\ are driven at higher rates for various 
reasons.
The carbon isotope exchange reactions also involve C\p\ and it is of
interest to ask whether they also must be driven at non-thermal rates
in order to account for the observed N(\cotw)/N(\coth) ratios.

The rate constant for thermal recombination of \hcop\ with electrons
is $3.3\times10^{-5}\ccc~\ps/\TK$ according to the UMIST reaction rate
database \citep{WooAgu+07}; therefore the volume formation rate of \cotw\
due to recombination of \hTWcop\ in a gas having an electron
fraction \Xe = n(e)/n(\HH) is 

\def\kf{\mbox{k$_{\rm f}$}}
\def\kr{\mbox{k$_{\rm r}$}}
\def\kHe{\mbox{k$_{\rm He}$}}


$$ {\rm dn}(\cotw)/dt =  
 6.6\times10^{-14} ~\Xe {\rm n}(\HH)^2 \pccc~\ps/\TK $$

and the formation rate  of \coth\ due to recombination of 
\hTHcop\ is here taken to be 60 times smaller following our
isotope measurements in mm-wave absorption \citep{LucLis98}.

\subsection{Carbon isotope exchange}

\begin{figure*}
\psfig{figure=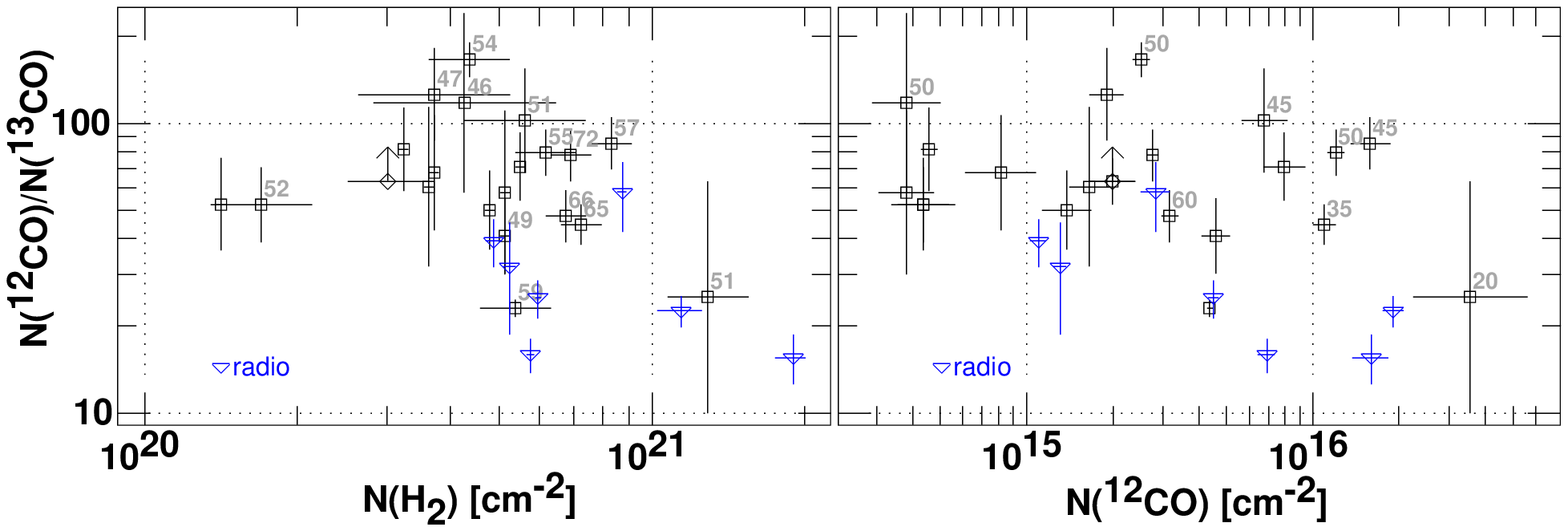,angle=-90,height=6cm}
\caption[]{Left: Variation of N(\cotw)/N(\coth) with N(\HH) (left) 
and N(\cotw).  Where possible points are labeled with estimates of \TK\ from
\HH\ (at left) and C$_2$ (right) as tabulated by \cite{BurFra+07} and 
\cite{SheRog+07} for \HH\ and \cite{SonWel+07} for \HH\ and C$_2$.  
Radiofrequency data are those of \cite{LisLuc98}.  There is no tendency
for the column density ratio to vary monotonically with the \HH\ temperature
and only a very loose trend with that of C$_2$}

\end{figure*}

\cotw\ molecules are interconverted to \coth\ with a  rate constant \kf\
by the reaction \THCp + \cotw\ $\rightarrow$ \TWCp + \coth\ + 34.8 K,
and in the other direction at rate \kr\ = \kf exp(-34.8/\TK) 
\citep{WatAni+76}.  The rate constants were measured by 
\cite{SmiAda80} and shown to be strongly temperature-dependent below 
500 K with a measured value $\kf = 7\times10^{-10}\ccc~\ps$ at \TK = 80 K,
(and an implied value 70\% larger at 10 K), as
compared with $\kf = 2\times 10^{-10}\ccc~\ps$ measured 
at \TK = 300 K by \cite{WatAni+76}. \cite{SmiAda80} discuss the means 
by which to convert their measurements to values of \kf\ below 80 K 
and their results have been employed by most subsequent authors, 
for instance \cite{ChuWat83} or \cite{LanGra+84} (although not by 
\cite{SheRog+07} who used the much smaller value of \cite{WatAni+76},
claiming that it agreed better with observation).
  
 A suitable expression for \kf(\TK) based on the 
results of \cite{SmiAda80}  does not exist in the literature.  For future 
reference, we provide the following:

$$ \kf = 7.64 \times 10^{-9} ~\TK^{-0.55} \ccc~\ps ~~(\TK = 80-500 {\rm K}) \eqno(4a) $$  

$$ \kf = {1.39 \times 10^{-9} ~\TK^{-0.05} \ccc~\ps 
 \over {1+\exp{(-34.8/\TK)}}}
  ~~( \TK = 10 - 80 {\rm K}) \eqno(4b) $$  

In terms of volume formation rates, the forward reaction forming \coth\ 
proceeds at a volume rate

$$ {\rm dn}(\coth)/{\rm dt}  = \kf ~{\rm X}(\cotw) ~{\rm X}(\THCp)
 ~{\rm n}(\HH)^2 \pccc~\ps $$

It is straightforward to show that the backward reaction between \TWCp\ and \coth\ 
is never an important ``source'' of \cotw\ (given the rate at which \cotw\ putatively
forms from H\cotw\p ) but conversion of \cotw\ to \coth\ is the dominant route
to \coth\ formation when the relative abundance X(\cotw) substantially exceeds 
$10^{-6}$.  The \THC-insertion reaction would dominate at much smaller X(\cotw)
if the direct formation rates were slower, leading to problems making models
with X(\cotw)/X(\coth) $>> 60$, as in Fig. 16  of \cite{WarBen+96} 
(at far left).

\subsection{Destruction of carbon monoxide by He\p}

The reaction He\p\ $+$ CO $\rightarrow$ C\p\ + O + He proceeds with a 
rate constant \kHe = $1.6\times 10^{-9}\ccc~\ps$ and is the dominant 
{\it chemical}
destruction mechanism for \cotw\ whenever n(\HH) $< 100 \pccc$.  {\footnote
The same is not true for \coth, whose chemical destruction is generally dominated by 
$^{12}$C-insertion whenever n(\HH) $ > 2\pccc$, see Eq. 4c and 4d.}
Calculation of the ionization equilibrium is complicated, but diffuse 
cloud models tend to produce a nearly constant $number~density$ of He\p\ 
 because such a large fraction of the free electrons arises 
from the near-complete photoionization of carbon.  From our models we 
take n(He\p) $= 3.4\times10^{-4}\pccc$ which is appropriate either when 
neutralization by small grains is considered and \zetaH\  
$= 2\times 10^{-16}~\ps$ per H-nucleon or when neutralization by small 
grains is ignored and \zetaH\ $=  10^{-17}~\ps$ \citep{Lis03}.  

The volume destruction rate of \cotw\ by He\p\ is 
 ${\rm dn}(\cotw)/{\rm dt} = -\kHe ~{\rm n(He)}\p {\rm n}
 (\cotw) \pccc~\ps$ or, numerically

$$ {\rm dn}(\cotw)/{\rm dt} = -5.4\times10^{-13} ~{\rm X}(\cotw)
 ~{\rm n}(\HH) \pccc~\ps $$

\subsection{Photodestruction and selective photodissociation}

The nominal free-space photodissociation rate  in the mean interstellar
radiation field is usually taken to be the same for either version of CO
and we parameterize the photodestruction rate internal to a cloud as 
 $  2.1\times 10^{-10} ~\ps ~{\rm g}_{\rm Y}  ~{\rm I}$
where I $<> 1$ represents the possibility of a variable external
radiation field, $ 2.1\times 10^{-10} ~\ps $ is the photodissociation
rate in free space \citep{LeTMil+00}  , 
and $ {\rm g}_{\rm 12}$ or $ {\rm g}_{\rm 13}$
represent the diminution of the photodestruction rate due to shielding 
by dust, \HH\ and other carbon monoxide molecules for either 
isotope.  The free-space rates for \cotw\ and \coth\ are taken
equal but $ {\rm g}_{\rm 12}$ declines more rapidly into a cloud than
does $ {\rm g}_{\rm 13}$ (\cite{vDiBla88,WarBen+96}). 
In Sect. 5 we show an example where  this behaviour is parametrized as   
$ {\rm g}_{\rm 13} = {\rm g}_{\rm 12}^{0.6}$.

\subsection{Relative abundances of \cotw\ and \coth}

Given the preceding considerations in this Section we may write the following
approximate, implicit and local expressions for the relative abundances 

$$ {\rm X}(\cotw) =  {{1.2\times10^{-7} {\rm X(e)}^\prime {\rm n}(\HH) /\TK}
\over { {\rm g}_{12} {\rm I} +0.0026+2.3\times10^{-5} 
\kf^\prime {\rm n}(\HH) }}   
\eqno(4c) $$

$$ {\rm X}(\coth) 
 = {({2.0\times10^{-9}{\rm X(e)}^\prime/\TK + 
 2.3 \times 10^{-5} {\rm X}(\cotw)){\rm n}(\HH)}
\over{  {\rm g}_{13} {\rm I~} +0.0026+0.0014 
\kf^\prime {\rm n}(\HH) \exp{(-34.8/\TK)}}}
 \eqno(4d) $$

Where $\kf^\prime = \kf/10^{-9}\ccc~\ps $, I measures the strength of the incident
interstellar radiation field normalized to its mean strength (I=1) and
X(e)$^\prime$ = X(e)$/4\times10^{-4}$ ; the electron fraction is comprised
of a contribution from fully once-ionized atomic carbon at the level of 
$3.2\times10^{-4}$  for an \HH\ gas  \citep{SofLau+04}, and the remainder from 
H\p\ due to cosmic-ray ionization \citep{Lis03}.  
The first term in the numerator of Eq. 4d corresponds to recombination
of H$^{13}$CO\p\ with X(H$^{12}$CO\p)/X(H$^{13}$CO\p) = 60 and the numerical
constant in both denominators represents destruction by interaction with
He\p .
In considering these
expressions note that the hydrogen in diffuse gas is not wholly molecular, 
probably even when CO is detected, and the total density n(H) = n(H I) + 2 n(\HH)
= 2 n(\HH)/\fH2\ where \fH2\ is the fraction of H-nuclei in \HH .

Note the following with regard to these expressions. For \cotw , 
photodestruction is dominant until g$_{12}~{\rm I} \approx 0.003$, at which 
point the shielding would be strong enough to permit near-complete 
conversion of carbon to CO.  Thus this simple chemical scheme can 
carry the gas from the diffuse to the dark regime (where
the free carbon abundance is somewhat lower and  X(\cotw) 
$\approx 10^{-4}$).  Also for \cotw, destruction by conversion to 
\coth\ dominates over destruction by He\p\ only for n(\HH) $> 100 \pcc$; this
 stands in opposition to the situation for \coth, where the interaction 
with He\p\ is negligible at almost all densities.  This imbalance contributes
to the lack of equilibration of the carbon isotope exchange, complicating
the interpretation of the observed \cotw/\coth\ ratios,

Finally, note that conversion from \cotw\ becomes the dominant source 
of \coth\ only when X(\cotw) $> 10^{-4}/\TK$ or 
X(\cotw) $> 2\times 10^{-6}$.
If the direct formation rate of carbon monoxide were taken to be much 
smaller than that given here, as is the case in models which 
substantially fail to reproduce the overall carbon monoxide 
abundance, conversion from \cotw\ would dominate the formation of 
\coth\ at far smaller X(\cotw).  This would make it very make it 
very difficult to reproduce ratios  N(\cotw)/N(\cotw) $>>$ 60,
which require strong self-shielding of the \cotw\ and therefore, 
high X(\cotw) (see Sect. 5.2).

\begin{figure}
\psfig{figure=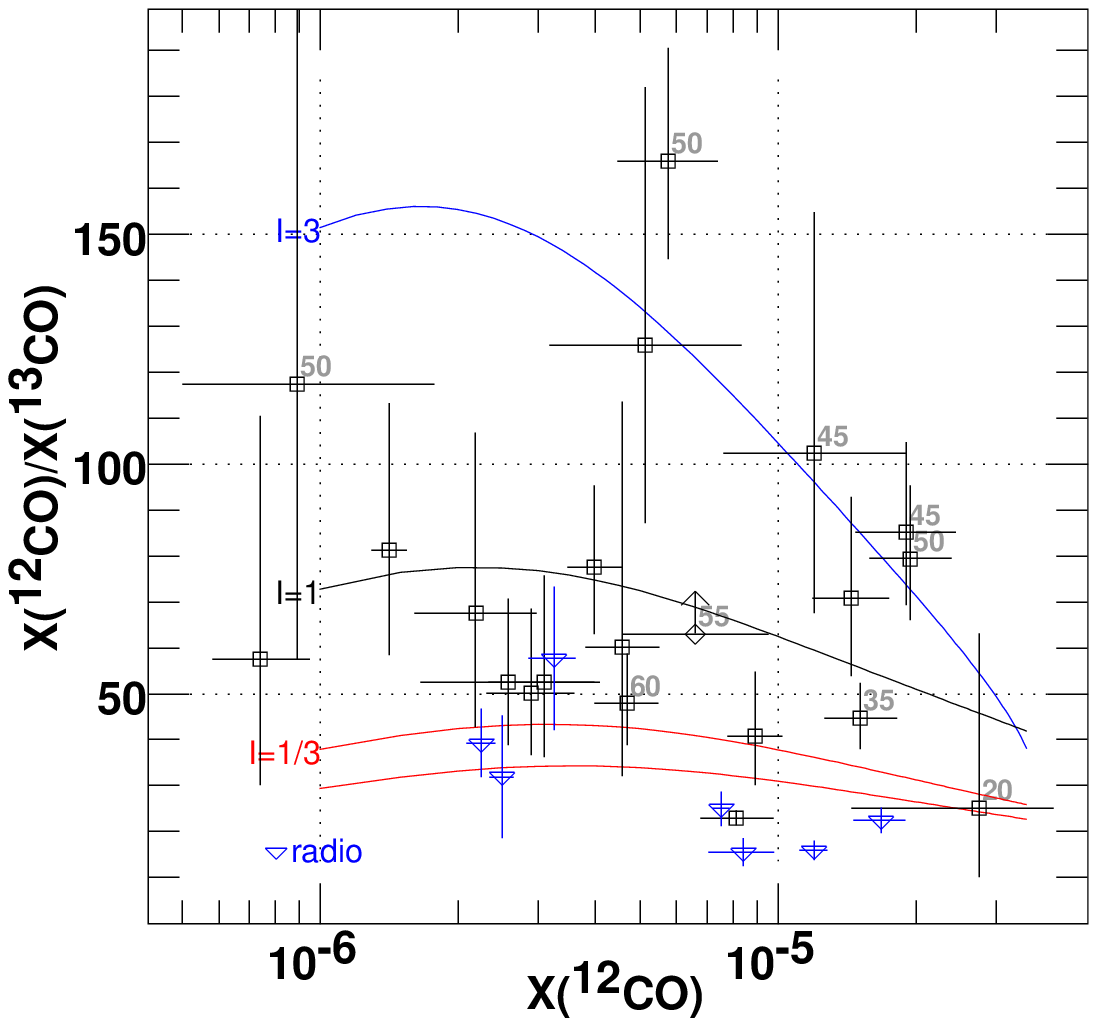,height=8cm}
\caption[]{Variation of N(\cotw)/N(\coth) with X(CO) = N(\cotw)/N(\HH) 
compared with the model chemistry (curves) described in Section 4.5 of the 
text. Models are for  
I=3, \TK=45 K, n(\HH)= 50 $~\pccc$;
I=1, \TK=45 K, n(\HH)=125 $\pccc$;
I=1/3,  \TK=22.5 K, n(\HH)=125 $\pccc$ (upper of two curves) 
and 250 $\pccc$.  As in Fig. 3 optical data are labelled with the kinetic 
temperature derived from C$_2$ as tabulated by \cite{SonWel+07}.
}
\end{figure}

\section{Observed N(\cotw) and N(\coth)}

\subsection{X(CO) at the extremes of abundance}

Figure 2 shows the observed variation of X(\cotw) with N(\HH); with much
scatter, the regression line has unit slope (see the caption); Fig. 20
of \cite{SonWel+07} appeared to show that X(\cotw) $\approx 5\times 10^{-8}$,
independent of N(\HH), for most lines of sight at 
N(\HH) $ < 3\times 10^{20}\pcc$.  

\subsubsection{The low-abundance limit}

Considering the smallest observed CO column densities and abundances,
at N(\HH) $ \le 3\times 10^{19}\pcc$, $<$X(\cotw)$>$ 
$ \approx 8\times 10^{-8}$ (Fig. 2).  In light of Eq. 4c, we 
note that n(H)/\TK\ $ \approx 1$ in diffuse gas, for instance, 
n(H) = 50$\pccc$, \TK= 50 K yields a typical thermal pressure in the 
ISM, but n(\HH)/n(H) $<$ 1 in such a gas.   Thus even our high 
{\it ad hoc} formation rate for CO predicts X(\cotw) somewhat below 
$10^{-7}$ in unshielded regions g$_{12} = 1$ and the observed CO 
is likely to be largely unshielded at the lowest observed N(\HH).
Gas observed at N(\HH) $> 10^{20}\pcc$, X(\cotw) $> 10^{-7}$ 
must generally be substantially
self-shielded in \cotw, implying that the observed N(\cotw)/N(\coth\
do not directly reflect the interstellar isotope ratio.

\subsubsection{High X(CO)}

Complete conversion of carbon to CO in a fully molecular diffuse gas 
would yield X(\cotw) $= 3.2 \times 10^{-4}$ given the free gas phase 
abundance of carbon determined by \cite{SofLau+04}, implying that 
the highest observed fractions of carbon in carbon monoxide in the
molecular portion of the gas are [\cotw]/[\TWC] $\approx 0.08$.  These
fractions are small enough to ensure that the observed gas may truly be 
considered diffuse, with nearly all carbon in the form of C\p, 
but they are not necessarily small enough that their
effects on other species are completely ignorable.

For \THC, the fraction  in \coth\ may be substantially higher owing to 
fractionation, reaching 0.20 - 0.25 (see just below).  In this case, the
ambient molecular gas is appreciably deprived of free \THC-nuclei, and 
account should be taken of such effects on the isotope ratios derived 
from other species like CH which are expected to share the same 
volume as CO.  Such sharing is somewhat less obvious for CH\p\ but in 
general, it seems clear that carbon isotope measurements in the diffuse 
ISM should avoid sightlines having substantial CO, unless the bias 
caused by carbon monoxide fractionation can be corrected somehow. 
It seems generally to be the case that the more pronounced effect of 
CO fractionation is to deprive the ambient gas of \THC\ but,
if it were observed that N(\cotw)/N(\coth) $>>$ 60 at sufficiently 
high X(\cotw), the opposite would be the case.

Again considering Eq. 4c, but in the limit of large X(CO), 
X(CO) $\rightarrow 4.7\times10^{-3}/$\TK\ in the limit of high n(\HH) 
and strong shielding ( I g$_{12} \rightarrow 0$) or
X(\cotw) $\approx 10^{-4}$ at \TK\ = 40 K.  Noting the behaviour 
in Fig. 1 and 2 we infer that this high-abundance limit would occur 
only for N(\HH) $ >> 3\times10^{21} \pcc$, 
N(\cotw) $\approx 3\times 10^{17}\pcc$, well above the observed
range.  X(\cotw) = $10^{-4}$ is still a factor of three below the free-carbon 
fraction relative to \HH\ in diffuse molecular gas. 
In the observed diffuse sightlines, the
 destruction of \cotw\ is dominated by photodissociation so that
 Ig$_{12} \ga 0.01$ even at the largest N(\HH) and N(\cotw).

\subsection{Fractionation}

Figure 3 shows a summary of available measurements of the two most abundant 
forms of carbon monoxide observed in absorption in diffuse clouds.  
\cite{SheRog+07} noted that lines of sight studied in \coth\ have
about 20 K lower \HH\ J=1-0 rotational temperatures than the mean for
all \HH\ surveyed.  As indicated in Fig. 2, the sightlines examined in 
\coth\ in {\it uv} absorption are nearly always those with 
the very highest X(\cotw) at a given N(\HH), and to that extent are not 
entirely representative even though they vary widely in column density.

\begin{figure*}
\psfig{figure=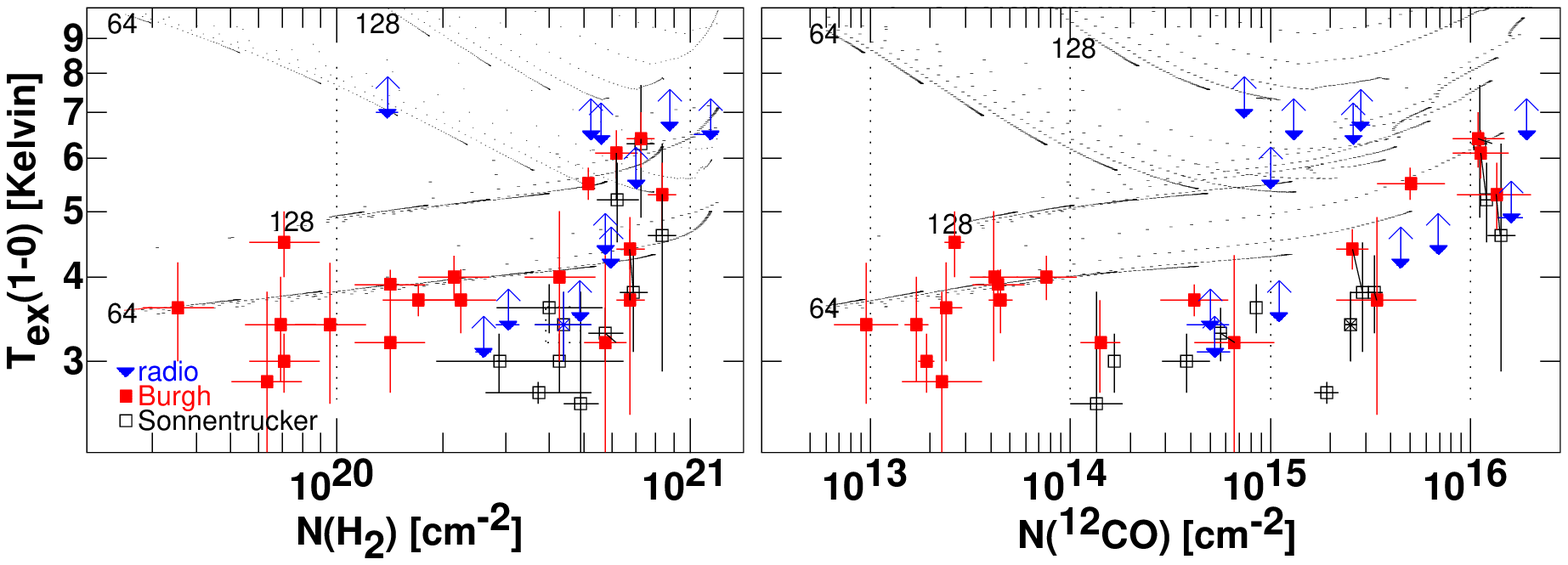,height=6.4cm}
\caption[]{Variation of  \cotw\ J=1-0 rotational excitation temperature
plotted against N(\HH) (left) and N(\cotw).   Radio data are shown
as lower limits because they involve the assumption of a beam efficiency.  
Data from the work of \cite{BurFra+07} and \cite{SonWel+07} 
are shown separately; sightlines with measurements in both references are 
shown chained, note that the former typically derives slightly stronger 
excitation in these cases.  Data labelled ``radio'' are those of 
\cite{LisLuc98}.  Superposed on the data are calculated results for 
uniform-density spherical models  having total densites n(H) = 64 and 
128 $\pccc$ like those used to determine N(CO) in Fig. 1, where only 
a handful of the sightlines represented here require n(H) $< 64 \pccc$.   
The lower curves at each density employ an older but likely more nearly 
correct set of rotational excitation rates for H-atom + CO collisions
(see Sect. 6 of the text).    A minimum excitation temperature of 4.0 K is
required to produce a $\lambda 2.6$mm emission line having a brightness 
temperature \TB = 1 K.}
\end{figure*}

The data show N(\cotw)/N(\coth) ratios in the range 
15 $<$ N(\cotw)/N(\coth) $<$ 170 and a tendency for the ratio 
to decline with increasing column density and/or relative abundance 
X(CO), but with much scatter.  There is no tendency for the ratio to
decline with the rotational temperature of \HH\ as noted in the
Figure but the two lines of sight with the smallest ratios studied
optically are those with very much the lowest temperature indicated by
C$_2$.   The radio data consistently find
smaller ratios at a given column density but the effect is more
pronounced in the panel at right.  There has been something of a 
recent convergence between the radio and {\it uv} absorption studies 
in that the latter have only now found lines of sight where 
N(\cotw)/N(\coth) $<<$ 60, as was often 
found to the case in mm-wavelength absorption \citep{LisLuc98}.  
Previously, optical absorption studies occured only along lines 
of sight having much larger ratios \citep{LamShe+94,FedLam+03} and 
this disparity was the source of some concern.

The expected functional behaviour of the column density ratio is as 
follows; at very low N(\HH) and/or N(CO) the gas is unshielded, 
the photodissociation rates are assumed equal and the observed 
N(\cotw)/N(\coth) ratios should be very near the ratio of formation
rates, presumably [\TWC]/[\THC] = 60 \citep{LucLis98}.  However,
as indicated in Fig. 2, the lines of sight chosen for study of the
isotope ratio have such high CO abundances that they do not sample
the unshielded regime.   At intermediate X(CO) and N(CO), self-
shielding increases the contribution of isotope exchange 
to \coth\ formation at X(\cotw) $> 2\times 10^{-6}$, but also increases 
the disparity between  g$_{12}$ and \g13 .  Therefore, at
intermediate X(CO) the column density ratio N(\cotw)/N(\coth) can be
expected to be both above and below the intrinsic isotope ratio
in the gas, with the very lowest values at larger N(CO).  Finally,
at very high N(CO), the column density ratio must tend toward the intrinsic
isotope ratio when all $^{12}$C resides in \cotw\ but this regime
again is well beyond the scope of the present dataset.

In order that the forward and backward isotope insertion reactions 
equilibrate,
the rightmost terms in the numerator and denominator of Eq. 4d must 
dominate.  In the numerator of Eq. 4d this implies  X(\cotw)  
well above $ 2\times 10^{-6}$  or that 
g$_{12} < 0.1$ in Eq. 4c; photodissociation remains the dominant
mechanism of CO destruction well into the regime where fractionation
is important.  For g$_{12}$I = 1/33, n(\HH) $= 40 \pccc, \TK = 40$ K, 
X(\cotw) $= 4\times10^{-6}$.  If the disparity between the isotopic
shielding terms is a factor of a few,  
 Ig$_{13} \approx 0.07-0.1$, and dominance of the righthand term 
in the denominator would require n(\HH) $>> 150~\pccc$.  Thus, 
although there is substantial creation of \coth\ $via$ isotope
exchange , it does not occur in a portion of parameter space where 
the isotope insertion actually equilibrates, creating a reliable gas
thermometer.

\subsection{What do we learn from \cotw/\coth\ ratios? }

Figure 4 shows the observed CO abundance ratios plotted
against X(\cotw) along with the results of some toy models
following the expressions in Eq. 4c and 4d, along with the
additional parametrization ${\rm g}_{13} =  {\rm g}_{12}^{0.6}$ .   
The parameter
having the greatest influence on the observed isotope ratio
is the external photodissociation rate, rather than the
density or temperature.  In detail, these observations are
very hard to interpret in terms of either the detailed physical 
properties of the gas or the intrinsic carbon isotope ratio.
The old hope that CO fractionation would serve as a reliable 
thermometer is not fulfilled.

However, in very general terms it does appear that 
the isotopic abundance ratios seen in carbon monoxide can 
be understood in terms of relatively mundane chemical
and photo-processes in quiescent gas of moderate density,
even if it is difficult to derive the underlying physical
conditions from the ratios themselves.  This is analogous
to the recognition that CO can form in the observed
quantities from the quiescent thermal gas phase electron 
recombination of the observed amounts of \hcop, and in very
great contrast to considerations of the abundances of \hcop\ 
and most other molecular species observed in diffuse gas, 
certainly all the polyatomics and many diatomics like CS,
whereby quiescent models fail by very large factors.


\section{J=1-0 rotational excitation and line brightness}


\subsection{Observed rotational excitation}

In optical absorption, carbon monoxide rotational excitation temperatures 
are derived directly from measurements of column densities in 
individual rotational levels \citep{BurFra+07,SonWel+07} and N(\HH)
is generally known.  At mm-wavelengths, the measurements
are of optical depths and line brightnesses in rotational transitions
\citep{LisLuc98} and the \HH\ column density is inferred 
indirectly, by assuming N(\hcop)/N(\HH) = $2 \times 10^{-9}$. 

Shown in Fig. 5 are J=1-0 rotational excitation temperatures \Texc(1-0)
from these references.  They are very nearly unanimous in showing (at left)
that lines 
of sight with N(\HH) $\la 5\times10^{20}\pcc$ have \Texc(1-0) $\la $ 4 K; 
4.0 K is the smallest excitation temperature capable of producing a 1.0 K
brightness temperature above the cosmic microwave background (CMB) in 
the $\lambda 2.6$mm J=1-0 line.  The same
data are plotted against N(\cotw) at right in Fig. 5, where some disparity
occurs between the radio and optical absorption data; a few lines of 
sight having N(\cotw) $\approx 10^{15}~\pcc$ exhibit somewhat brighter 
J=1-0 lines than would be allowed by a 4 K excitation temperature.
 
Figure 6 shows the observations recast as integrated $\lambda 2.6$mm
J=1-0 rotational brightness temperatures 
\WCO\ = $\int \TB(1-0) {\rm dv}$.  
At radio wavelengths, these are directly observed
to within a scale factor, the beam efficiency.  For optical
absorption data, the line brightness can be crafted from
the rotational level populations, which fix the integrated optical
depth; the central optical depth is then specified by the b-
parameter, so that the line brightness can be integrated over 
the profile.  For optically thin lines the integrated 
brightness is independent of the b-value, but many of the lines 
of sight with N(\cotw) $> 10^{15}\pcc$ observed in {\it uv} absorption
are predicted to be somewhat opaque at $\lambda 2.6$mm, 
in keeping with direct measurements shown by \cite{LisLuc98}.  
For \Texc(1-0) = 2.73 K or 4.0 K, the integrated optical depth of 
the J=1-0 transition is 1 \kms\ when N(\cotw) = 1.0 or 
$1.6 \times 10^{15}\pcc$.  Finally, note that \cite{BurFra+07} 
published only N(\cotw) and \Texc(1-0); to specify the rotational 
level populations fully we assumed that the excitation 
temperatures of higher-lying levels were the same as that of 
the J=1-0 line.  The excitation is weak enough that this is not
a major problem, little population exists in levels above J=2.

At left, Fig. 6 shows the integrated brightness temperature and 
\HH\ column density for the optical absorption data.  Also shown
are the predicted \citep{SonWel+07} and observed \citep{Lis97}
brightnesses toward \zoph ; the excitation temperature derived
by \cite{SonWel+07} is noticeably smaller than earlier values
\citep{SmiSte+79,WanPen+82} leading to a low predicated \WCO\ and 
larger implied optical depth in the J=1-0 line.  Corresponding
data comparing WCO and N(\hcop) were not provided by  \cite{LucLis98}, 
so, to test the consistency of the optical and mm-wave results, 
as elsewhere in this work, we used a set of unpublished, very 
sensitive, low galactic latitude CO emission and \hcop\ absorption 
profiles and simply compared the integrated CO brightness and 
\hcop\ optical depths, scaling the latter appropriately to form 
N(\hcop) and N(\HH) as done by \cite{LucLis98} and \cite{LucLis96}.  
This is the one exception noted in Sect. 2 to the use of 
previously-published data.  In any case, it can be seen from 
Fig. 6 at left that the radio and optical data show the same 
behaviour.

This comparison of \WCO\ and N(\HH) is tantamount to specifying 
the CO-\HH\ conversion factor; indeed, this regime is really
the only one in which the conversion factor may be derived from
direct measurement of the two constituents.  Clearly the ratio of
N(\HH)/\WCO\ is typically very large for weak-lined diffuse 
CO-bearing gas, compared with the usually assumed local values 
N(\HH)/\WCO\ = $2-3\times10^{20} \HH/$K-\kms.  However, for sightlines 
of sight with \WCO\ $>$ 1 K \kms, the ratio is nearly canonical or 
perhaps a bit low.  This insight, that the CO-\HH\ conversion factor
attains nearly its canonical values at rather small N(CO) was at the 
heart of the original discussion of \cite{Lis82}.

\begin{figure*}
\psfig{figure=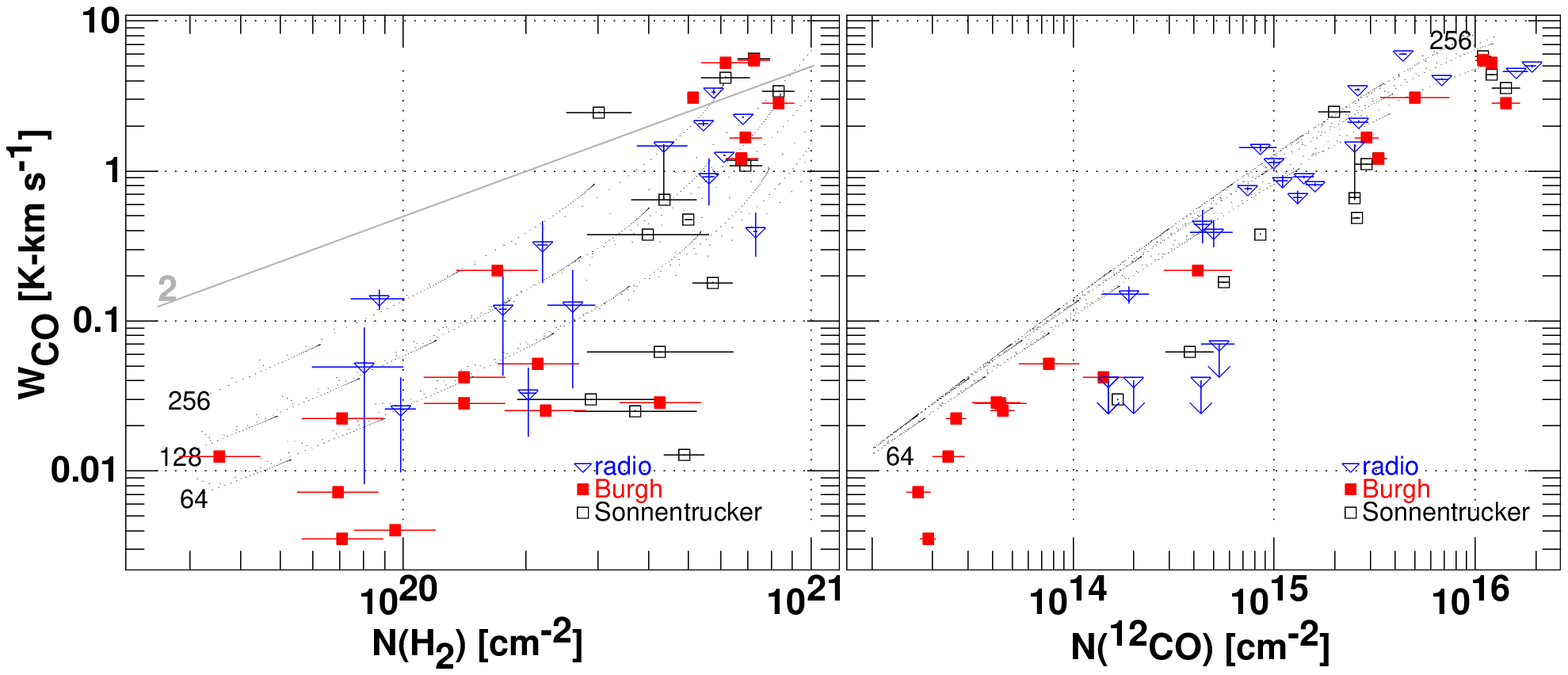,angle=-90,height=7.5cm}
\caption[]{ Integrated \cotw\ J=1-0 $\lambda 2.6$mm brightness temperatures 
predicted from {\it uv} absorption and/or observed at $\lambda 2.6$mm, plotted 
against N(\HH) (left) and N(\cotw) (right).  The shaded line at left 
represents a CO-\HH\ conversion factor  N(\HH) = $2 \times 10^{20}\pcc$/K-\kms .  
Direct radio observations toward \zoph\ are also shown chained to the
equivalent optical datapoint.  Data from the work
of \cite{BurFra+07} and \cite{SonWel+07} are shown separately and the
radio observations are those of \citep{LisLuc98} at right; see Sect. 6.1
for an explanation of the radio data plotted at left.}
\end{figure*}

The plot of WCO $vs.$ N(\cotw) at right in Fig. 6
shows a well-defined proportionality, as noted earlier (Fig. 12
of \cite{LisLuc98}) and there is good agreement between the optical
and radio data when CO emission was actually detected at $\lambda$
2.6mm, at N(\cotw) $> 6 \times 10^{14}\pcc$. Apparently, \WCO\ is a 
fairly robust and accurate estimator of the CO column density itself
and scatter in the CO-\HH\ conversion factor arises from the vagaries
of the chemistry. Note that Fig. 6 at right reverts to use of the 
previously published mm-wave data. 

For N(\cotw) $\approx 3-5 \times 10^{15}\pcc$ the brightness saturates 
at \WCO\ $\approx 3-6$ K \kms\ as the lines become very optically 
thick.  This highlights the fact that the transition to complete 
C$\rightarrow$CO conversion must occur over a relatively small range 
of N(\HH); if diffuse cloud lines of sight commonly show J=1-0 lines 
of several K, not grossly different from typical dark gas even when 
only a few percent of the free carbon is in carbon monoxide, and 
if the CO-\HH\ conversion factor is about the same for dark and 
brighter-lined diffuse gas, it follows that there must be a very 
rapid increase of X(CO) over a very small interval in N(\HH) and 
\WCO , over and above that seen in Fig. 1.

\subsection{Interpretation of the rotational excitation}

Until recently, it seemed appropriate to neglect the contribution
of excitation by H-atoms, for which the cross-sections were calculated
to be relatively small \citep{GreTha76}, to assume a largely 
molecular host gas, and
to use the CO J=1-0 rotational excitation temperature as a probe of 
the ambient thermal pressure; it could be shown from calculations 
that there was a nearly linear increase of \Texc(1-0) with 
n(\HH)\TK\ \citep{SmiSte+79,Lis79,LisLuc98} as long as 
\Texc\ $<<$ \TK.  This variation of \Texc(1-0) with the ambient thermal
pressure of \HH\ presumably occurs as a result of the small 
permanent dipole moment of CO since the excitation of other species with 
higher dipole moments is typically sensitive mostly to density.
 However, a recent calculation of the rotational excitation rates 
of CO by H-atoms yielded the result that the per-particle
excitation by hydrogen atoms was  much stronger than by \HH\
\citep{BalYan+02}.  In this case even a small admixture of H-atoms can 
have a marked effect on the excitation of CO \citep{Lis06} and it is 
not allowable to derive the ambient thermal partial pressure 
of \HH\ directly from \Texc(1-0) by assuming a purely molecular 
host gas.

 As this manuscript was being revised, it was learned (P. Stancil, 
private communication, \cite{SheYan+07}) that the calculation 
of \cite{BalYan+02} had been reconsidered and that the older, 
smaller H-atom excitation rates were in fact more nearly appropriate.  
Results for both sets of rates are shown here.  The choice of excitation 
rates has several interesting consequences for the interpretation, as 
noted below.

Model results for the CO excitation are superposed In Fig. 5 and 6.  These
were calculated by running models ( like those used to calculate N(CO)
in Fig. 1) at fixed n(H) (as indicated in the Figures) and N(H) toward 
the center, plotting results from sightlines having impact parameters
ranging from the outer edge to center.  For each model of given n(H) and 
N(H), the rotational level populations were integrated for each sightline 
and the result plotted as a point in the Figure.  Results for n(H)
below 64 $\pccc$ are not shown because only a handful of the 
sightlines represented in Fig. 5 are compatible with n(H) $< 64 \pccc$
in Fig. 1 (where they are shown outlined and in green to make just this
point). 

 Shown in 
Fig. 5 are results for models having n(H) = 64 and 128 H-nuclei $\pccc$
using the H-atom excitation rates of \cite{GreTha76} (as 
approximated by \cite{WarBen+96}) and those of \cite{BalYan+02}.
The models include excitation by atomic H and He, and ortho and para-\HH
and the effects of photon-trapping were calculated in the microturbulent 
approximation for a b-parameter of 0.8 \kms.  The characteristic shape
of the upper three curves in Fig. 5 is determined by an increase in 
the fraction of atomic H toward the left, and by resonant 
photon-trapping at higher optical depth toward the right. With strong
excitation by atomic hydrogen, calculated values of \Texc(1-0) increase 
with n(H) at small N(\HH) or N(\cotw).   Models using
the older cross-sections for excitation by atomic hydrogen do not exhibit
such an increase because the overall excitation is weak when the molecular
fraction is small.


 Only if the smaller H-atom excitation rates are appropriate will 
the models used to calculate X(CO) also do an acceptable job of
reproducing the observed excitation at like density.
The excitation temperatures derived from the {\it uv} absorption data
are a bit lower than before, but mostly there is the problem that
the fraction of gas which remains in atomic form in the models
(which self-consistently determine the local \HH\ density) is 
large enough to have a profound effect on the excitation.  Lines
of sight with the smallest N(\HH) and X(\cotw) are expected to 
have formed CO in the least purely molecular gas and therefore 
to show the highest excitation temperatures, opposite to what is 
observed  , if the rates of \cite{BalYan+02} are employed.
Except perhaps for the anomalous-seeming radio data near N(\cotw) 
$= 10^{15}\pcc$ at the right in Fig. 5, the  calculated rotational 
excitation is far too high, mainly due to the admixture of 
residual atomic hydrogen in the CO-bearing regions.

The situation with respect to the integrated brightness shown in Fig. 6 
is  clearer, as the models actually account quite well for the 
variation of \WCO\ with N(\HH) at left and they only slightly 
overestimate the run of \WCO\ with N(\cotw).  Furthermore, the 
calculated values of \WCO\ are not strongly dependent on which 
set of cross-sections for excitation by H atoms is used.  This 
implies that it may be possible to calculate the CO-\HH\ conversion 
factor reliably, although it should be stressed that the models
form CO rather artificially.  Moreover, if the larger H-atom 
excitation rates of \cite{BalYan+02} are employed, the disparity in 
the ability of the models
to account for both \Texc\ and \WCO\ implies that the calculated
optical depths are much lower than those which are actually observed 
at $\lambda 2.6$mm or inferred from the {\it uv} absorption data.

\section{Loose ends and unanswered questions}

The comparison of CO with \HH\ in diffuse clouds can now be based on a 
dataset which far exceeds that for any other trace molecule.  Over the 
diffuse regime, the 
run of N(CO) with N(\HH) actually seems more consistent and understandable 
than that of CH, which suffers from an unexplained decline in
its relative abundance at N(\HH) $< 7 \times 10^{19}\pcc$ (see Fig. 1);
at higher N(\HH), N(CH)/N(\HH) $\approx 4\times 10^{-8}$ with relatively
small scatter.  Further observations of CH and other species are required 
in the regime of moderate N(\HH) and nascent polyatomic chemistry.

On average, the mean N(\cotw)/N(\coth) ratio is near the local interstellar
isotope ratio [\TWC]/[\THC] = 60, but fractionation effects cause
the N(\cotw)/N(\coth) to vary between 20 and 170 with lower values at somewhat
higher N(\cotw).  The fraction of free gas-phase \THC\ in \coth\ is large 
enough in some cases that other species sharing the same volume must be
somewhat starved for \THC , artificially biasing the ratio of their 
\TWC- and \THC-bearing variants.  However, the effects of 
sharply-varying fractionation might perhaps be most pernicious for 
mm-wave emission studies, which typically rely on ratios of brightness 
temperatures in \cotw\ and \coth, coupled with the assumption of a fixed 
abundance ratio N(\cotw)/N(\coth), to infer \cotw\ optical depths, 
excitation temperatures and column densities.  Unrecognizable systematic
variations in  N(\cotw)/N(\coth) could wreak havoc with interpretation of
such datasets.

At the present time it is unclear whether fractionation in CO can be 
regarded as anything more than a nuisance: the observed N(\cotw)/N(\coth)
ratios result from several competing influences including direct formation,
carbon isotope exchange and selective photodissociation, none of which 
dominates to the extent that the local temperature or density can be 
inferred.  At least in the {\it uv} datasets, \cotw/\coth\ ratios have generally 
been measured along lines of sight having only rather high X(CO) and over a 
relatively modest range of N(\HH) (see Fig. 2).  This may have introduced 
some bias but interpreting the observations is 
sufficiently difficult that further discussion of the fractionation may 
not be rewarding {\it per se}.  However, study of \coth\ may be 
rewarding for other reasons, especially at higher N(\cotw) and N(\HH) 
as more complete carbon conversion to CO occurs.

The convergence of the CO-\HH\ conversion factor for diffuse and dark
gas well within the diffuse regime (at N(CO) $\la 10^{16}\pcc$ in Fig.
6 at left) suggests that the full conversion of carbon to CO must occur
over a very narrow range of extinction, N(\HH), $etc.$ just beyond
the diffuse regime.  This may leave very little room in parameter space
for a ``translucent'' regime in which neutral carbon is the dominant
carbon-bearing species.  Given that the CO abundance in dark gas is
typically found to be X(CO) = $ 8-10 \times 10^{-5}$, some 3-4 times
smaller than  2[C]/[H] $\approx 3.2 \times 10^{-4}$ in diffuse gas, we
are left to ask just where the transition from diffuse to dark gas 
actually occurs in terms of N(\HH) and what is  the carbon budget 
in the transition regime.

In the diffuse regime, the CO-\HH\ conversion factor is actually 
measured and found to attain values 
N(\HH)/\WCO\ $= 2-3 \times 10^{20} \HH/$K-\kms\ along rather thin lines
of sight where the fraction of free gas-phase carbon in CO is only a
few percent.  These lines of sight also have \WCO\ of a few K-\kms,
as is typical of darker material.  Again we are left to wonder how much
room in parameter space is actually left for a translucent regime.

The observed CO-\HH\ conversion factors and CO J=1-0 rotational line
brightnesses are actually well-explained by the uniform density models 
whose results are presented in Figs. 1, 5 and 6.  The most notable 
 possible failure in the interpretation here is the mismatch between 
the too-large excitation temperatures and too-small optical depths predicted 
by the models for the J=1-0 rotational transition (see Fig. 5 at left)
when the recent calculation of the H-atom + CO excitation rates by 
\cite{BalYan+02} is employed (see Sect. 6.2).  The models are uniform and 
rather diffuse and so leave a substantial fraction of H-nuclei in atomic 
form  in some cases.  Excitation by such H atoms was ignorable using 
older excitation rates \citep{GreTha76} 
 but the rates of \cite{BalYan+02} are so large that even a slight amount of
residual atomic hydrogen would have a profound effect.  Although the disagreement 
between observed and measured excitation temperatures seems to point
to the need to sequester CO in regions of nearly pure \HH , the agreement 
is actually worst for the most diffuse gas with the smallest N(CO) and X(CO).
Should such gas really be expected to be the most purely molecular ? 
 Perhaps the neatest way around this problem lies with the apparent 
recent realization that the older, smaller excitation rates by H atoms are 
actually more nearly correct.  In this case the models do a good job of 
reproducing the rotational excitation of CO at the same densities at 
which CO forms, even if the ambient hydrogen is not exclusively molecular.

There is little difference in \cotw\ J=1-0 line brightness between 
dark clouds with 
X(\cotw) $ = 10^{-4}$, N(\cotw) $ = 3-10\times 10^{17} \pcc$
and that from diffuse clouds with 
X(\cotw) $ = 2\times 10^{-5}$, N(\cotw) $ = 3-10\times 10^{15} \pcc$.
Although dark and diffuse gas may better be distinguished on the 
basis of observations of \coth\ (acknowledging possible effects of
fractionation) or \coei, such data are not always available.  
Given the prevalence of diffuse gas at larger distances from 
the center of the Galaxy and larger distances from the galactic plane, it
seems worthwhile to ask whether possible confusion between diffuse and dark
gas has caused misjudgement of the quantity and character of molecular 
gas in our own or other galaxies.  

\begin{acknowledgements}

The National Radio Astronomy Observatory is operated by Associated 
Universites, Inc. under a cooperative agreement with the US National 
Science Foundation.  I am grateful to Phillip Stancil for informing
me of the very recent calculations of CO excitation by H atoms due to
\cite{SheYan+07} and to Malcolm Walmsley for a host of worthwhile
comments.  The finishing touches were put on this paper while the 
author was enjoying the hospitality of IRAM and the Hotel Hesperia 
in Granada.

\end{acknowledgements}
 

\end{document}